\documentclass{aastex}
\usepackage{emulateapj5}
\usepackage{epsfig}

\newcommand{\thalf}{t_{1\!/2}}
\newcommand{\tE}{t_E}
\newcommand{\tthalf}{\tilde{t}_{1\!/2}}
\newcommand{\ttE}{\tilde{t}_E}

\begin{document}
\title{Event rates and timescale distributions from \\
realistic microlensing models of M31}
\author{Edward A. Baltz$^1$, Geza Gyuk$^{2,3}$ and Arlin Crotts$^1$}
\affil{1.\ ISCAP, Columbia Astrophysics Laboratory, 550 W 120th St., Mail Code
5247, New York, NY 10027\\
2.\ Department of Astronomy and Astrophysics, University of Chicago, Chicago,
IL 60637\\
3.\ Adler Planetarium and Astronomy Museum, Chicago, IL 60605}

\begin{abstract}
We provide a set of microlensing event rate maps for M31, the Andromeda Galaxy.
Rates for M31 microlensing were calculated on the basis of a four component
model of the lens and source populations: disk and bulge sources lensed by
bulge, disk, M31 halo and Galactic halo lenses.  We confirm the high rate
gradient along the minor axis of M31 due to a dark halo of lenses.
Furthermore, we compute the timescale distributions of events, for both
Einstein times and full-width at half-maximum times.  We explore how the rate
contours and the timescale distributions can be used to measure the shape and
extent of the microlensing halo.  With one year of twice--weekly sampling, or
three observing seasons, a halo MACHO fraction as small as 5\% can be detected
with modest ground based telescopes.
\end{abstract}


\section{Introduction}
Gravitational microlensing as a means to detect compact objects in the Galactic
halo was first considered by Paczy\'nski (1986), but the basic idea is much
older (Einstein 1936).
This suggestion was realized as results from two surveys of microlensing events
towards the Large and Small Magellanic Clouds (LMC and SMC) (Alcock et
al.~2000; Ansari et al.~1996).  These were consistent with a significant, but
subdominant, contribution of microlensing masses to the Galactic dark matter
halo.
Nonetheless, these conclusions are still controversial, and the identity and
location of the microlensing masses are still mysterious.

A decade ago, M31 was suggested as a promising venue where galactic
microlensing might be explored in ways advantageous and distinctive from that
in and around the Galaxy (Crotts 1992).  Several papers (Jetzer 1994; Han \&
Gould 1996; Baltz \& Silk 2000; Kerins et al.\ 2001) have confirmed that a
substantial microlensing signal can be expected.  Two collaborations, MEGA
(preceded by the VATT/Columbia survey) and AGAPE have produced a number of
microlensing event candidates involving stars in M31 (Crotts \& Tomaney 1996,
Ansari et al.~1999, Auri\`ere et al.~2001, Uglesich 2001, Calchi Novati et
al.~2002).  Here we show the potential for these and future surveys to settle
some of the outstanding questions regarding microlensing in spiral galaxies.

This is the second paper in a series.  Paper I (Gyuk \& Crotts 2000) provides
optical depth maps for M31. While these are useful tools for certain purposes
they are unfortunately not directly measurable. Event rates on the other hand
{\em are} directly measurable and hence their magnitude and variation across
the face of M31 are more meaningful in planning and evaluating surveys of
microlensing.
In this paper we extend Paper I to include event rate maps, both total and also
differential rates with respect to the event timescale.

This paper is organized as follows.  In \S\ref{model} we briefly discuss the
M31 models we used, including disk, bulge and halo components.  Following this
we present rate maps for various halo models, including the self lensing
contribution, in \S\ref{ratemaps}.  In \S\ref{timescales} we provide
differential rate distributions as a function of two different timescale
measures, and we discuss how cuts in timescale can be used to separate self
lensing from a MACHO halo contribution to the lensing rate.  We conclude with a
discussion of measuring lens masses and halo properties in \S\ref{discussion}.

\section{Modeling}
\label{model}

The execution of a microlensing survey of M31 is qualitatively different from
those towards the Magellanic Clouds or our own Galactic bulge.  For the latter,
individual stars can often be resolved, and their unlensed fluxes measured,
admitting a direct measurement of the magnification during a microlensing
event.  In contrast, M31 is more than ten times farther away, and individual
stars are rarely resolved from the ground.  Images are almost always highly
crowded, and individual sources are almost always highly blended.  This is the
``pixel lensing'' regime (Crotts 1992; Baillon et al. 1993).  Image
subtraction, necessitated by the high degree of crowding, has been very
successful.  While this technique allows the detection of variable objects of
any kind, and is photon noise limited, the unlensed fluxes of lensed sources
are unknown.  Thus M31 microlensing surveys must make do with a full width at
half maximum timescale and a flux increase at maximum as the useful fit
parameters.  The unlensed flux measured by classical microlensing surveys is
unavailable under most circumstances.

Sources are taken to reside in a luminous two-component model of M31 consisting
of a double exponential disk and a bulge.  The disk model is inclined at an
angle of 77$^\circ$ and has a scale radius of 5.98 kpc, a scale height of 400
pc, a central surface brightness of $\mu_R = 20$, $V-R=0.67$, and we ignore
spiral arm effects (Walterbos \& Kennicutt 1988; Gould 1994).  We take an $I$
band surface brightness profile from the $V-R$ color and the observation that
$V-R\approx R-I$ (Bessel 1979).  The rotation velocity is taken to be 240 km
s$^{-1}$, with a linear rise from the center to 6 kpc.  The bulge model is
based on the ``small bulge'' of Kent (1989) with a central surface brightness
of $\mu_R = 14$, and $R-I=0.72$ (Walterbos \& Kennicutt 1988; Bessel
1979). This is an axisymmetric bulge with a roughly $\exp(-r^{0.4})$ falloff in
volume density with an effective radius of approximately 1 kpc and axis ratio,
$c/a \sim 0.8$. Values of the bulge density are normalized to make $M_{\rm
bulge} = 4 \times 10^{10} M_\odot$.  The velocity distribution of bulge stars
is taken to be maxwellian ($dN\propto\exp(-v^2/2\sigma^2)d^3\vec{v}$), with
\mbox{$\sigma=150$ km s$^{-1}$}.  These quantities are fairly well known, and
unlikely to change the results by a large amount if revised.

We explore a parameterized set of M31 halo models. Each model halo is an
axisymmetric cored ``isothermal sphere'' determined by two parameters: the axis
ratio or flattening $q$ (where $q=1$ indicates no flattening) and the core
radius $r_c$.  As we are concerned with lensing objects we also define the
MACHO fraction ($f_b$) as the fraction of the halo mass that consists of
lensing objects.  Together, the mass density of halo lensing objects is then
\begin{equation}
\rho(x,y,z) = f_b\,\frac{V_c(\infty)^2}{4 \pi G}
\frac{e/(q\,\sin^{-1}e)}{x^2+y^2 + (z/q)^2 +r_c^2},
\end{equation}
where $e=\sqrt{1-q^2}$ and $V_c(\infty)=240$ km s$^{-1}$ is taken from
observations of the M31 disk.  The velocity distribution of the halo is taken
to be maxwellian, with a circular velocity equal to $V_c(\infty)$, making
$\sigma=170$ km s$^{-1}$.

We compute rate distributions with respect to two different timescales.  The
first is the more familiar Einstein time $(\tE)$, defined as the time to cross
the full Einstein disk which has radius
\begin{equation}
R_E=\sqrt{\frac{4GM_{\rm lens}D_s\,x\,(1-x)}{c^2}},
\end{equation}
where $x$ is the fractional distance to the lens in relation to the distance to
the source, $D_s$.  The Einstein time is then given in terms of the lens
velocity perpendicular to the line of sight,
\begin{equation}
\tE=\frac{2R_E}{v_{\perp}}\,.
\end{equation}
The second timescale we use is the full--width at half maximum of the
lightcurve $\thalf$, much more easily measured when the source fields are
crowded, as is the case for M31.  This timescale is simply related to the
Einstein time and to the minimum impact parameter $\beta$ of the lens relative
to the line of sight (taken in units of $R_E$).  This timescale is given by
(Gondolo 1999)
\begin{equation}
\thalf=\tE w(\beta),
\end{equation}
with the following definitions
\begin{equation}
w(\beta)=\sqrt{2f(f(\beta^2))-\beta^2},\;\;\;
f(x)=\frac{2+x}{\sqrt{x(4+x)}}-1.
\end{equation}
The function $w(\beta)\sim\beta$ for all values of $\beta$, with the limiting
behavior
\begin{equation}
w(\beta\ll1)\approx\beta\sqrt{3},\;\;\;
w(\beta\gg1)\approx\beta\sqrt{\sqrt{2}-1}.
\end{equation}
Thus, $\thalf\sim\beta\tE$, and is more degenerate than $\tE$ due to the
dependence on impact parameter.  A further complication is that $\thalf$ is
essentially the measured timescale.  Determining the Einstein time requires
knowledge of the magnification, which can be very difficult in highly crowded
fields as the source stars are highly blended.  The Einstein time might be
inferred with extra effort, either by high resolution imaging of the source
star or statistically with the $t_\sigma$ technique (Baltz \& Silk 2000).  It
is crucial to evaluate if these are necessary.

\section{Rate maps}
\label{ratemaps}

Using the model of \S\ref{model}, we compute the rate of detectable
microlensing events.  Two computer codes written entirely independently and
described elsewhere (Baltz \& Silk 2000; Gyuk \& Crotts 2000) have been used,
and produce nearly identical results.  A numerical integration of the rate is
performed, over the positions and velocities of the source (with fixed
brightness) and lens (with fixed mass) along a given line of sight.  The
probability that an event with given parameters is detected is folded into the
the integral.  Mass functions for the lenses and luminosity functions for the
sources are applied at the end.

With lines of sight spaced at one arcminute intervals, we construct contour
maps of the event rate for ``self lensing'' by stars (assuming no binary
lenses, at most a 15\% contribution (Mao \& Paczy\'nski 1991; Baltz \& Gondolo
2001)) as well as for halo lensing.  The self lensing contribution has four
logical contributions, taking sources and lenses from the bulge and disk.  The
bulge-bulge contribution dominates the self lensing near the center of the
bulge, along the minor axis the bulge-disk and disk-bulge contributions
dominate, and far from the bulge the disk-disk contribution dominates (though
this is always dominated by the halo contribution for $f_b>0.05$).  These four
in sum give a self lensing rate that is nearly symmetric about both the major
and minor axes of M31.  The halo lensing contributions arise from both the M31
and Milky Way halos, lensing both disk and bulge stars.  The Milky Way
contribution has a nearly uniform optical depth.  The M31 contribution is
strongly asymmetric, with a significantly larger rate from the far edge of the
minor axis.

We have assumed the following definition for a detectable microlensing event.
The MEGA survey most frequently employs the MDM 2.4m telescope, thus we use its
capabilities in the following.  We assume that integrations totaling three
hours are taken twice weekly during the M31 observing season: this is fairly
conservative.  We define an event as a deviation that has two consecutive
samples four standard deviations above the baseline.  We assume one arcsecond
seeing, which is typical: the median MDM seeing is 0.95 arcsecond.  To
approximate the sensitivity of MDM, we assume that a star of $R$ magnitude
$m_R=25.2$ or $I$ magnitude $m_I=24.8$ gives one photoelectron per second, and
furthermore that the noise in the images is twice the photon counting noise
(this is quite conservative).  We take the distance modulus to M31 as $D=24.5$,
a distance of 795 kpc.  We assume a sky brightness of $\mu_R=21$ and $\mu_I=20$
mag arcsec$^{-2}$.  Here we reiterate the fact that the source stars are not
resolved, but instead are typically {\em highly} blended.  Only with difference
imaging can the microlens variability be detected.  The only measured
``baseline flux'' is the light falling within a resolution element due to
several stars in the highly crowded fields.

The luminosity function of M31 sources must be specified.  For the disk, we
take $R$ and $I$ band luminosity functions from Mamon \& Soneira (1982).  For
the bulge, we take the $I$ band from Terndrup, Frogel \& Whitford (1990).  Data
for the $R$ band is scarce, so we average (in $\log dN/dM$) $V$ and $I$ band
data for $M_R>0$ (using Terndrup et al.~(1990) for $I$ and Holtzman et
al.~(1998) for $V$).  For $M_R<0$ we attach a power law slope of 0.59 taken
from the MACHO project data (Alves 2001).

The mass function of lenses, both stellar and the MACHO component, is somewhat
problematic.  For the stellar component, we use the exponential Chabrier (2001)
mass function, down to $0.01 M_\odot$.  This is steeply decreasing, so there is
little mass in the lowest decade (the brown dwarfs).  Varying the stellar mass
function in acceptable ranges has little effect unless for example there is a
large component of brown dwarfs (Baltz \& Silk 2001).  The mean stellar mass of
about $0.5 M_\odot$ is what dominates the rate and timescale from self lensing.
We have also investigated a Scalo (1986) mass function, which has a slightly
higher mean mass, thus longer timescales and lower rates.  The differences are
not very large however.  For lack of convincing evidence to the contrary, we
take a fixed mass for MACHOs.  Based on Alcock et al.\ (2000) we assume a mass
of $-3/8$ dex relative to solar $(\approx 0.422 M_\odot)$, and we test values
of $0.1 M_\odot$ and $1.0 M_\odot$ as well.  We note here that MACHO mass can
have a large effect on the expected timescales and rates.  Larger masses would
indicate a lower rate and longer events, while smaller masses give a higher
rate but shorter events.  This is discussed further in the next section.

\begin{figure}
\epsfig{file=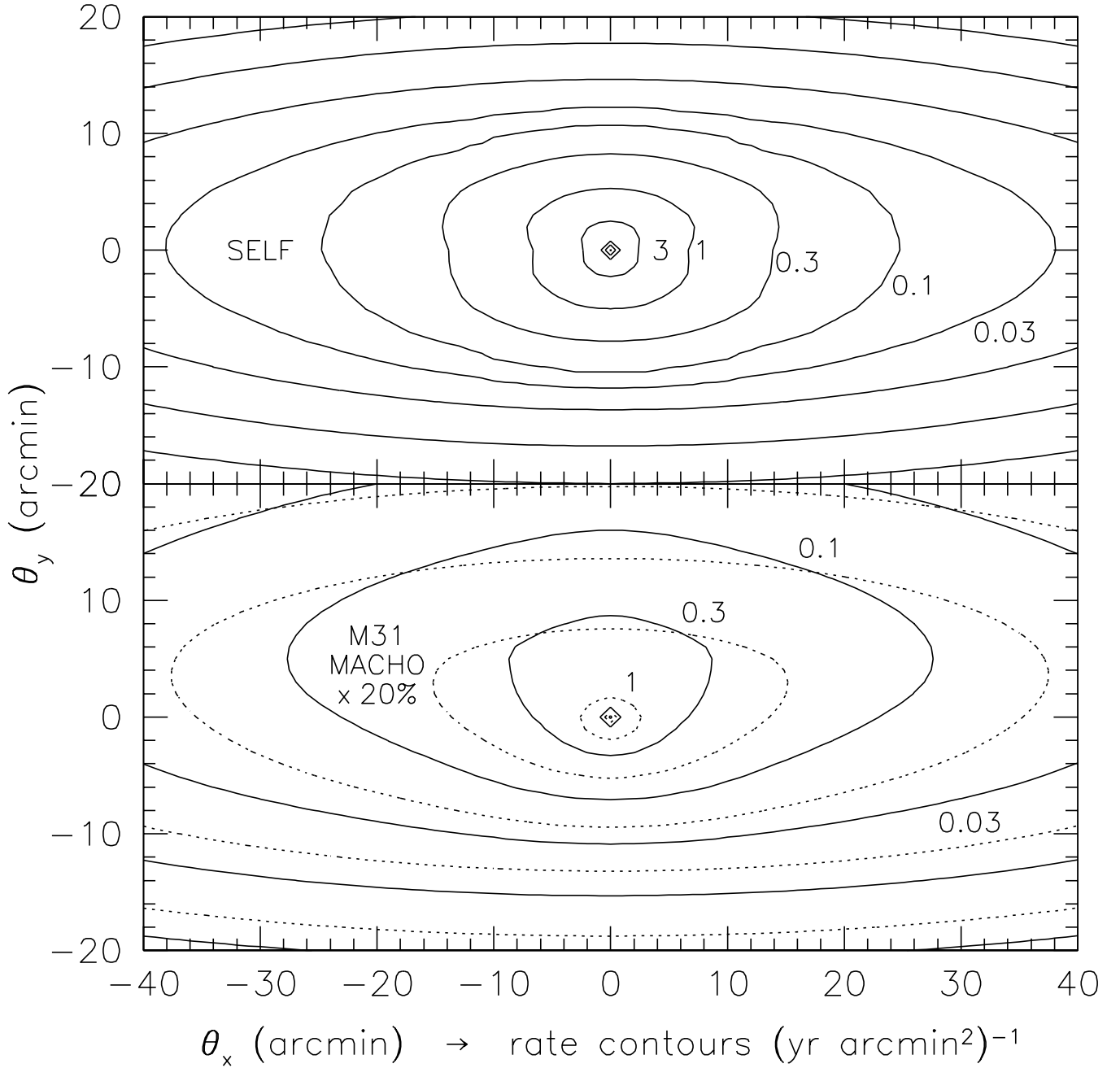,width=\columnwidth}
\caption{Rate contours for self lensing and M31 MACHO halos, in units of events
per year per square arcminute.  The top panel shows the self lensing component
only, while the bottom panel shows the contribution of halo lensing with
$f_b=0.2$ and $r_c=2$ kpc.  The solid contours indicate a round ($q=1$) halo
for M31, while the dotted contours indicate a flattened ($q=0.3$) halo.  The
number ``3'' in the contour labels should be taken as $\sqrt{10}$, i.e. the
contours are separated by 0.5 dex in units of \mbox{events yr$^{-1}$
arcmin$^{-2}$}.  The rate contours for self lensing are basically symmetric
from front to back, since the disk provides an asymmetrical source population
for bulge lenses, but also an asymmetrical lens population for bulge sources.
We see the strong asymmetry in the rate of MACHO events.}
\label{fig:map1}
\end{figure}

\begin{figure}
\epsfig{file=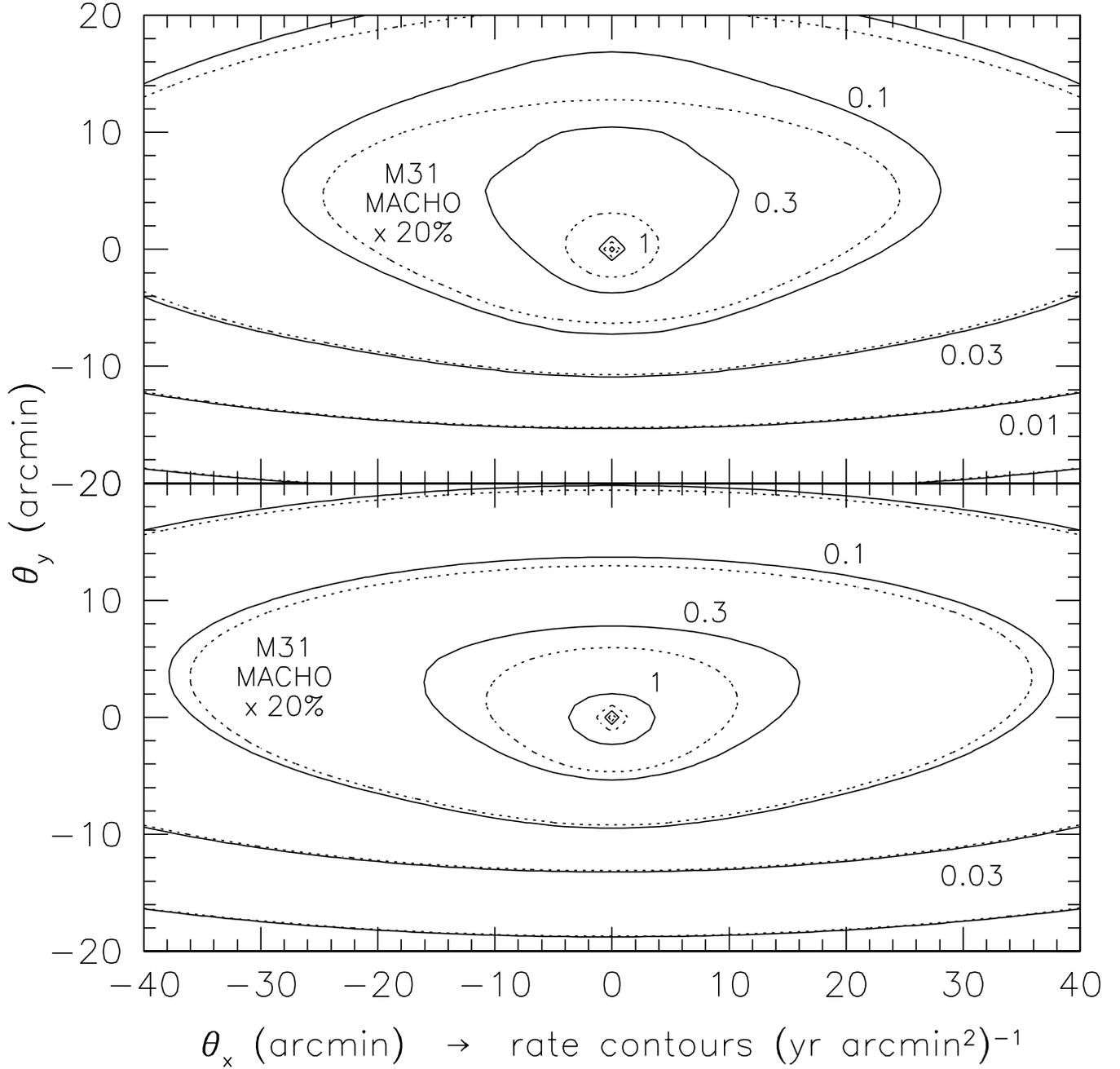,width=\columnwidth}
\caption{Rate contours for M31 MACHO halos, in units of events per year per
square arcminute.  All halos are taken with $f_b=0.2$.  The top panel shows
halo lensing for a round ($q=1$) halo, while the bottom panel shows the
contribution of a flattened ($q=0.3$) halo.  The solid contours indicate
$r_c=1$ kpc, while the dotted contours indicate $r_c=5$ kpc.  The labels are as
Fig.~\ref{fig:map1}.  It is evident that measuring the core radius from
microlensing will be difficult, especially in the case of a flattened halo.}
\label{fig:map2}
\end{figure}

\begin{figure}
\epsfig{file=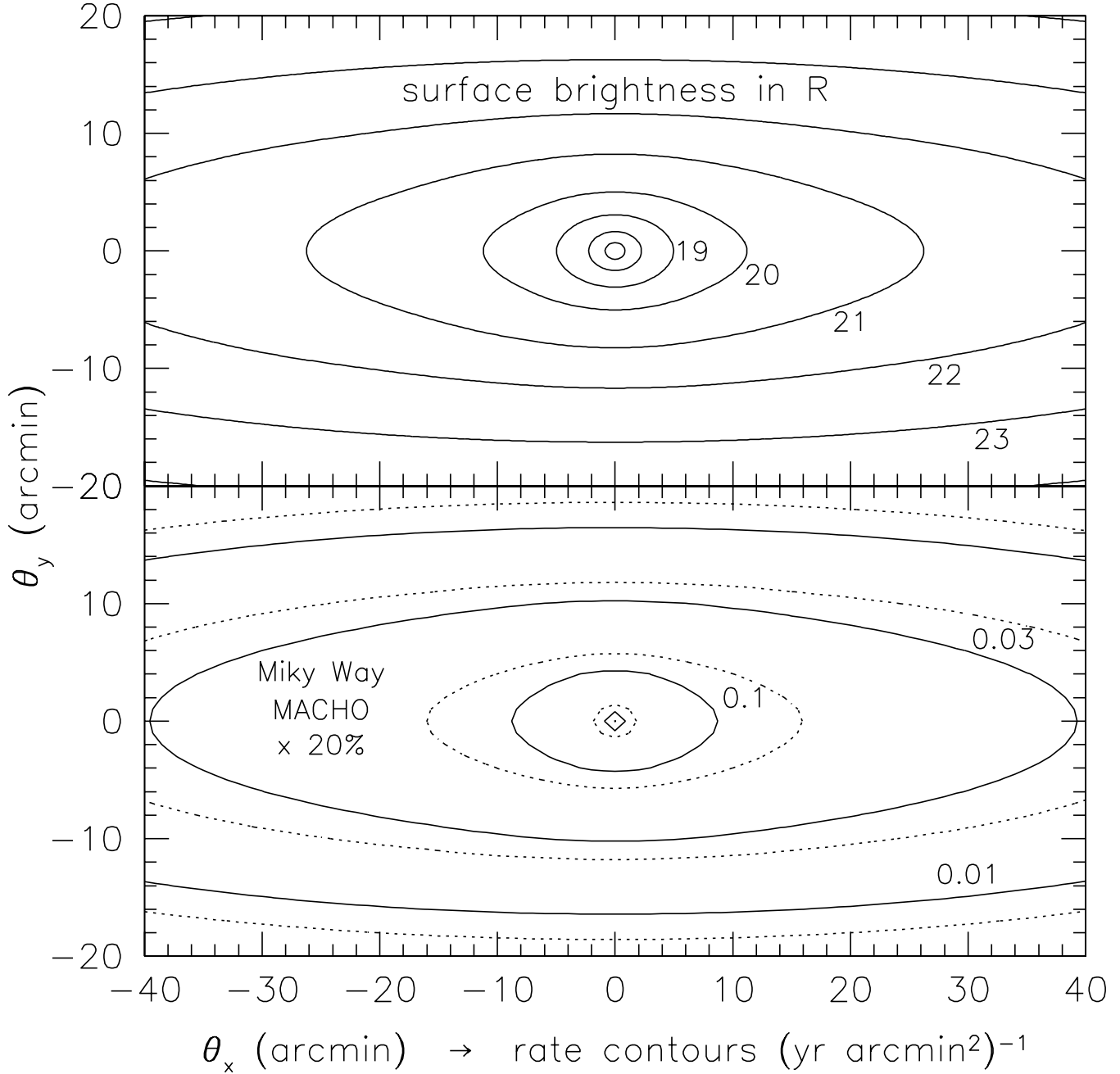,width=\columnwidth}
\caption{Surface brightness and rates for Milky Way MACHOs.  The surface
brightness is given in $R$ magnitudes per square arcsecond.  The Milky Way
MACHO rates are given in units of events per year per square arcminute.  The
halos are taken with $f_b=0.2$.  The top panel illustrates the surface
brightness contours.  The bottom panel illustrates the rates, with a $q=1$ halo
in solid contours and a flattened $q=0.3$ halo in dotted contours.  The labels
are as Fig.~\ref{fig:map1}.  The optical depth to Milky Way lensing is quite
uniform across M31, thus the rate only tracks the surface brightness, as
expected.}
\label{fig:map3}
\end{figure}

\begin{figure}
\epsfig{file=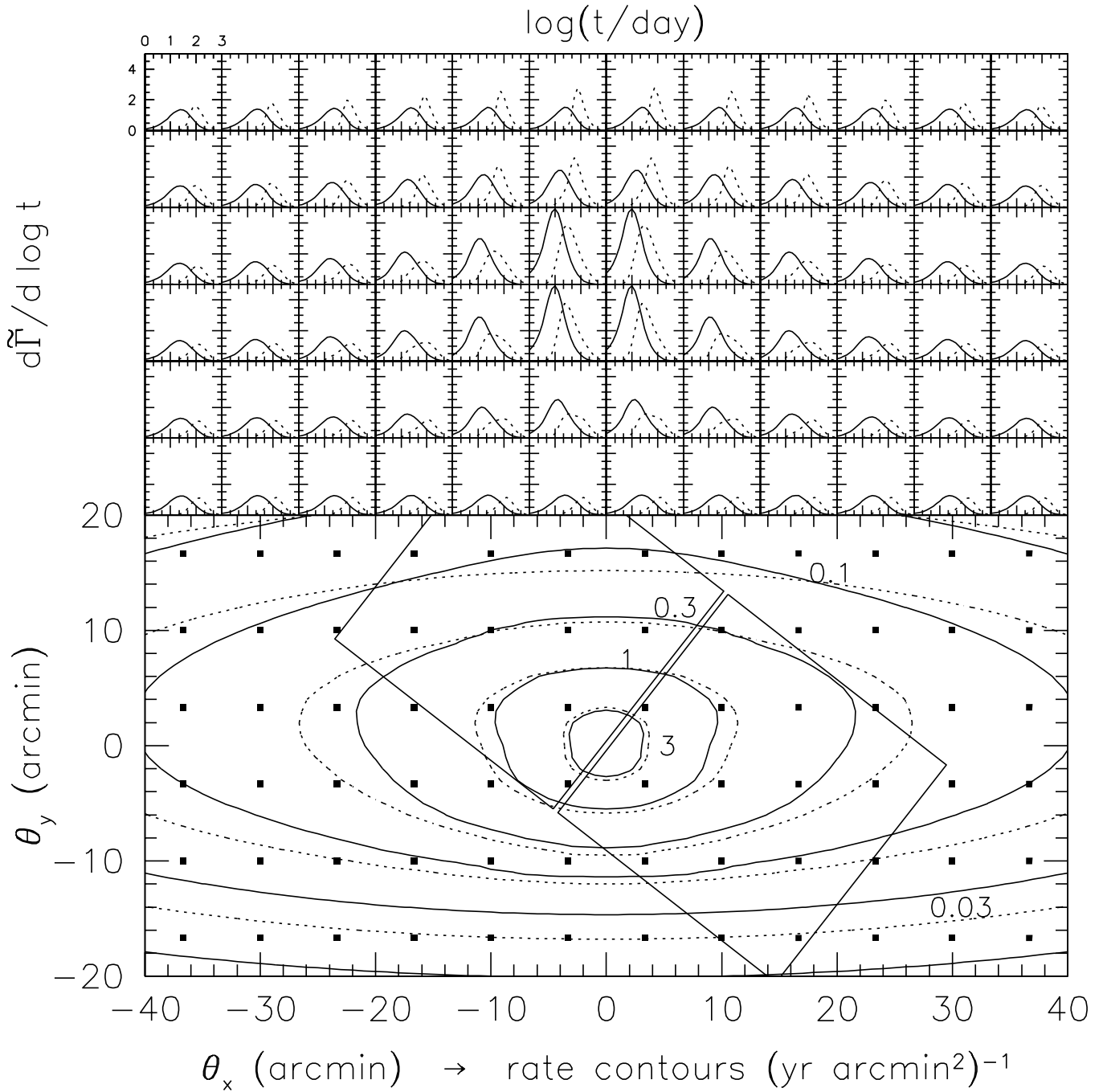,width=\columnwidth}
\caption{Rate contours for microlensing for different halo flattening values
(in units of events per year per square arcminute) and rate distributions for
two timescale measures.  The bottom panel illustrates the $f_b=0.2$, $r_c=2$
kpc case, with solid contours indicating $q=1$, and dotted contours indicating
$q=0.3$.  The self lensing and Milky Way contributions are included.  Self
lensing is dominant through most of the central region (with 5 arcmin of the
center).  The square MDM fields are illustrated.  The dots on the contour plot
indicate the lines of sight where we illustrate the distribution of timescales
$d\tilde{\Gamma}/d\log t$ in the small top panels.  The solid curves are for
$\thalf$, while the dotted curves are for $\tE$, both for the $q=1$ halo,
though the differences between that and the $q=0.3$ case are slight.  The
$\thalf$ plots are normalized to unity at $\log(\tthalf/{\rm day})=1.75$ (about
56 days), while the $\tE$ plots are normalized at $\log(\ttE/{\rm day})=2.25$
(about 178 days).  We thus see the variation in the long timescale tail across
the image.  In the central region the differential rate drops more quickly.}
\label{fig:map_dist}
\end{figure}

\begin{figure}
\epsfig{file=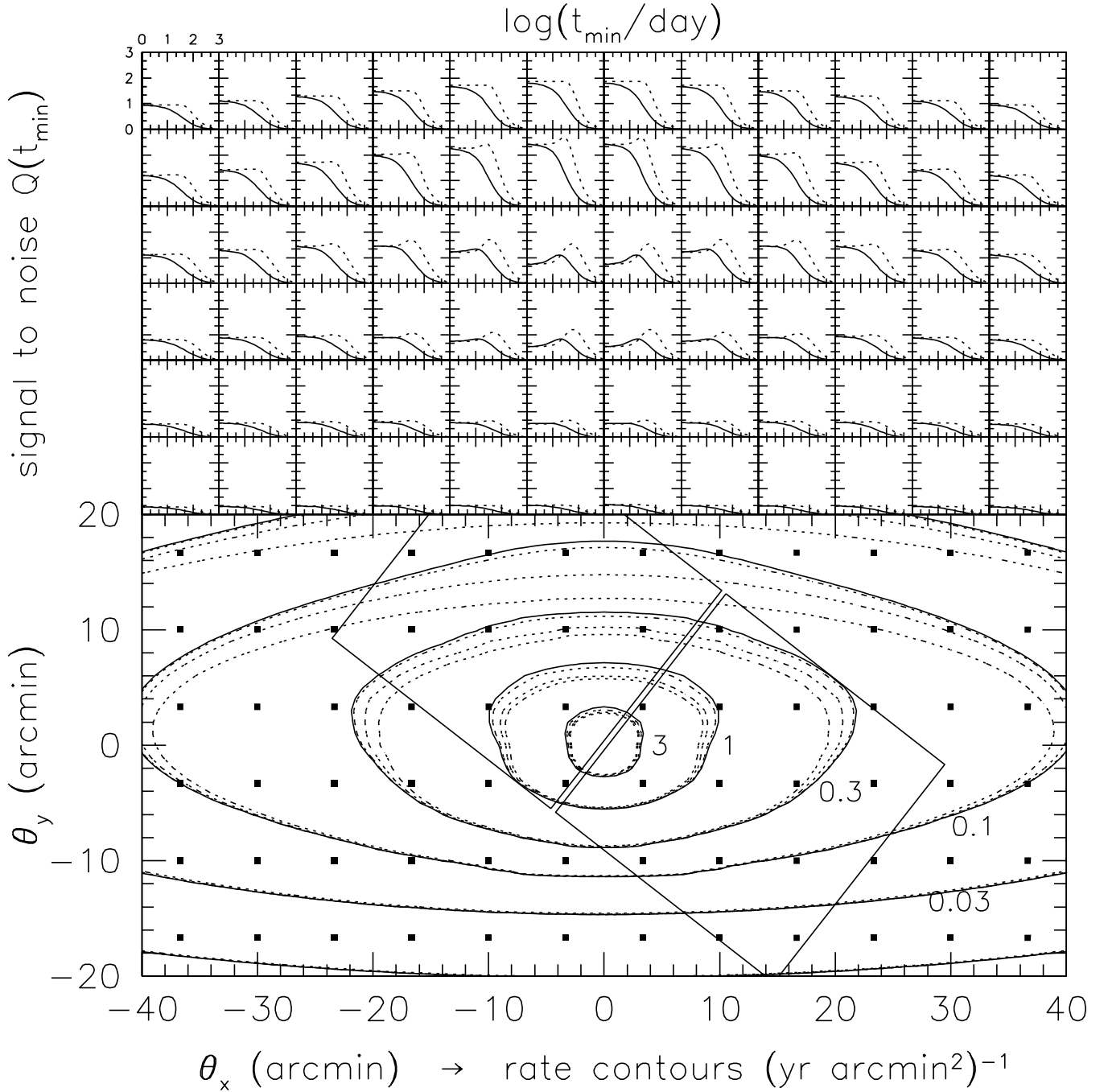,width=\columnwidth}
\caption{Rate contours for microlensing for different halo core radii (in units
of events per year per square arcminute) and signal to noise for separating
halo lensing from self lensing.  The bottom panel illustrates the $f_b=0.2$,
$q=1$ case, with solid contours indicating $r_c=1$ kpc, and the dotted contours
indicate, moving inward, $r_c=2,5,10$ kpc respectively.  There is a significant
difference in rate towards the far end of the minor axis as the core radius
varies, with the near end unaffected for the most part.  The self lensing and
Milky Way contributions are included.  In the small top panels we illustrate
the signal to noise (from Eq.~\ref{eq:sn}) for separating a halo from self
lensing for one year (three seasons) of observations over the box (about 45
arcmin$^2$), taking a timescale cut at $t_{\rm min}$.  Solid curves illustrate
the cut in $\thalf$, while the dotted curves illustrate the cut in $\tE$.  The
$\thalf$ cut doesn't help much, but if $\tE$ is known, it is quite advantageous
to make a cut near the center of the bulge.}
\label{fig:map_sn}
\end{figure}

In Figs.~\ref{fig:map1}-\ref{fig:map_mass} we plot event rate contours for $R$
band observations.  The contours for $I$ band are very similar.  The only
pronounced difference is in the bulge, where the $I$ band luminosity function
is quite shallow, there is a significantly larger rate in the most central 4
square arcminutes.  While it is desirable to detect events in two bands to test
that the flux {\em enhancement} has a constant color (in time), as should be
the case for gravitational lensing, we will not include this criterion.  The
separate event rates are very similar, thus we deem it unlikely that the joint
event rate will be much different.  In Fig.~\ref{fig:map1} we plot the rate
contours for self lensing and for M31 halos with a 20\% MACHO fraction c.f.\
Alcock et al. (2000) ($f_b=0.2$) and a core radius of 2 kpc.  Both round
($q=1$) and flattened ($q=0.3$) halos are illustrated.  These maps are given as
contours of constant rate, in units of events yr$^{-1}$ arcmin$^{-2}$.  In
Fig.~\ref{fig:map2} we plot rate contours again for $f_b=0.2$, but compare the
small ($r_c=1$ kpc) and large ($r_c=5$ kpc) core cases, for both round and
flattened halos.  In Fig.~\ref{fig:map3} we illustrate the event rate from
Milky Way MACHOs, assuming both round and flattened cases again.  The rate
should be related to the surface brightness, also illustrated.  In fact a naive
calculation indicates that the rate from a constant optical depth of lenses
(such as a Milky Way population, which wouldn't vary much over the M31 fields)
should be proportional to the square root of the surface brightness.  This can
be seen from the following simple argument.  The number of monitored sources is
proportional to the surface brightness (the proportionality constant depends on
luminosity function), but the noise level increases as the square root of the
surface brightness, necessitating an increase in magnification by the same
factor to obtain an equivalent signal to noise.  Since the peak magnification
of the event is proportional to the inverse of the impact parameter, the total
cross section goes down by this square root of surface brightness.  Thus, the
number of monitored sources times the cross section is proportional to the
square root of the surface brightness.  This argument breaks down at very high
surface brightness where the magnification required implies events whose full
width at half maximum timescales are too short to detect, however, this regime
is not reached for M31 for the masses assumed.  From Fig.~\ref{fig:map3} we see
that this calculation is reasonable, as the shape of rate contours matches the
shape of the surface brightness contours, and the rate of decline is roughly
half on the logarithmic scale illustrated.  The flattening of the Milky Way
halo only changes the normalization of the rates, not the shape of the
contours.

Combining all of these components, we illustrate the total expected event rate.
In the bottom panel of Fig.~\ref{fig:map_dist}, we plot contours of the total
microlensing rate, from self lensing and 20\% MACHO halos for both the Milky
Way and M31.  The self lensing dominates in the inner 5 kpc, but outside, the
20\% halo provides most of the events.  Changing the halo fraction $f_b$ has
little effect on the shape of the contours, and only in the region where the
self lensing and halo lensing are comparable (roughly speaking within the 0.3
events arcmin$^{-2}$ yr$^{-1}$ contour assuming a 20\% halo).  The overall
normalization of the event rate away from the bulge can give a clear
measurement of $f_b$.  For example, comparing a 20\% halo to the case of a 5\%
halo, the inner contours (within 1 event arcmin$^{-2}$ yr$^{-1}$) are not much
affected since self lensing dominates and the {\em shape} of the outer contours
is very similar since it is the halo that dominates there in both cases.
Between the 1.0 and 0.3 contours the event rate drops more quickly in the 5\%
halo case since this is the region where the microlensing rate from the halo
and stellar components are comparable.  This is clear from inspecting the top
panel of Fig.~\ref{fig:map1} (the self lensing rate) and comparing to the
bottom panel (the M31 halo lensing rate) which is trivially rescaled according
to MACHO fraction.

Fig.~\ref{fig:map_dist} shows the difference in the rate contours for round
$q=1$ and flattened $q=0.3$ halos.  The two cases appear to be easily
distinguishable, especially along the major axis.  In the bottom panel of
Fig.~\ref{fig:map_sn}, we again plot contours of the total microlensing rate,
this time varying the halo core radius from $1,2,5,10$ kpc.  The significant
difference is on the far side, along the minor axis.  A crude measurement of
core radius may thus be possible.

Here we see the utility of using the raw event rates to determine parameters.
We do not need to evaluate the optical depth to do so.  Since in the end we are
interested in the halo parameters like MACHO fraction and flattening, the
optical depth is secondary.

\section{Timescale distributions}
\label{timescales}

The Einstein time is the fundamental timescale parameterizing the variability
due to lensing, but it is notoriously difficult to measure from a pixel
microlensing event.  The directly measured parameter is the full-width at
half-maximum timescale.  We will illustrate the rate distribution of events in
both of these timescales in Fig.~\ref{fig:map_dist}, and we will discuss the
extraction of halo parameters using both.  The calculation of the differential
rate with respect to $\thalf$ follows Baltz \& Silk (2000), based on Griest
(1991) who discusses the differential rate with respect to $\tE$.  We plot the
differential rate with respect to the logarithm of the timescale, normalized by
the differential rate at a fixed timescale $\tilde{t}$,
\begin{equation}
\frac{d\tilde{\Gamma}}{d\,\log\,t}=\frac{d\Gamma}{d\,\log\,t}\;\Big/
\frac{d\Gamma}{d\,\log\,t}(\tilde{t}).
\end{equation}
Taking $\tilde{t}$ longer than the the peak timescale then illustrates the
variation in how rapidly the differential event rate falls at long timescales.
In the upper panels of Fig.~\ref{fig:map_dist} we illustrate
$d\tilde{\Gamma}/d\log t$ for both $\tE$ and $\thalf$.  For lines of sight
far from the bulge, the distribution in $\thalf$ peaks at around 20 days, while
the distribution of $\tE$ peaks at around 100 days.  Near the bulge, the peaks
of both timescale distributions are shorter.

Away from the bulge, where MACHO events dominate, the peak timescale is a
measurement of the MACHO mass, as $\tE\propto\sqrt{M_{\rm lens}}$.  Taking a
single mass, the $\tE$ distributions are $\approx0.5$ dex wide in $\tE$,
translating to $\approx1.0$ dex wide in $M_{\rm lens}$.  With 50 events, a
$\pm20\%$ measurement of the mass would be possible with known Einstein times.
With only $\thalf$, the situation is considerably worse, as the distribution in
$\thalf$ for a fixed mass is $\approx1.0$ dex wide in $\thalf$, thus $\approx
2.0$ dex wide in $M_{\rm lens}$.  With the same 50 events, a mass measurement
at the level of $\pm40\%$ can be made.  We thus see the utility of determining
the Einstein times for the events.

We expect MACHO events to have longer timescales than stellar events simply due
to the geometry, i.e.\ the source--lens distances can be much larger.  To
compare, the bulge is of order 1 kpc in size, while a cored halo does not drop
appreciably in density for of order 10 kpc.  Halo lenses have a larger velocity
dispersion than bulge lenses, but not by a large factor.  In all, we might
expect halo lens timescales to be a factor of 2.5 or so larger.  This fact can
be used to determine cuts in timescale, either $\tE$ or $\thalf$, that
maximizes the possibility that MACHO events can be distinguished from stellar
events.  We want a cut that both excludes as many self lensing events as
possible and includes as many MACHO events as possible.  We thus wish to
maximize the signal to noise, including both the Poisson counting of events,
and a systematic error due to the uncertainty in the self lensing model.  For a
given area, observation time and minimum timescale, we expect $N_{\rm M31}$
events from M31 MACHOs, $N_{\rm MW}$ events from Milky Way MACHOs, and finally
$N_{\rm S}$ events from stellar lenses, with an uncertainty of $\Delta N_{\rm
S}$ (which we take to be $0.3N_{\rm S}$).  We minimize the quantity
\begin{equation}
\frac{1}{Q^2}=\frac{N_{\rm M31}+N_{\rm MW}+N_{\rm S}}{N_{\rm M31}^2}+
\left(\frac{\Delta N_{\rm S}}{N_{\rm M31}}\right)^2
\label{eq:sn}
\end{equation}
as a function of minimum timescale to determine what, if any, timescale cut to
use.  These results are plotted in the top panels of Fig.~\ref{fig:map_sn}.
For most of the field, we find that no timescale cut should be made.  Near the
center of the bulge, within 6 arcmin on the minor axis and 12 arcmin on the
major axis, the situation is different.  If the Einstein times are known, it is
advantageous to cut all events with $\tE<75$ days.  This allows the largest
signal to noise for separating the halo component.  Without the cut, for a one
year survey the separation can be done at the $2.5\sigma$ level over the
interior region.  With the cut, the significance rises to 4$\sigma$.  If only
$\thalf$ is known, a similar timescale cut does not help much.  We comment here
that if we were to take a more conservative value for the model uncertainty
(say $\Delta N_S=0.5N_S$), the results are only affected near the center of the
bulge, but the desired cutoff in Einstein time and the signal to noise for
detecting a halo knowing the Einstein times are hardly affected even there.  We
also note that the utility of timescale cuts may change if the MACHO mass were
to be significantly different from typical stellar masses.  As it is now, the
MACHO project mass value and the typical stellar mass are quite similar.

Over the full far--side MDM field, a microlensing halo for M31 can be detected
at roughly $7\sigma$ significance with one full year of well--sampled data, or
approximately three observing seasons.  This assumes $f_b=20\%$, but the
significance is roughly linear with $f_b$.  For $f_b=10\%$, a microlensing halo
can be detected at roughly $4\sigma$ significance, and with $f_b=5\%$, the
significance drops to about $2.5\sigma$.  Thus, $f_b=5\%$ is about the limit of
detectability for a microlensing halo with three seasons of observing.
This 5\% value is also a plausible lower limit to halo lensing, caused by a
much greater number of stars in M31's halo/spheroid than the Galaxy's (Reitzel
et al.~1998, scaled to produce a microlensing prediction using Alcock et
al.~1997, Table 10).

\begin{figure}
\epsfig{file=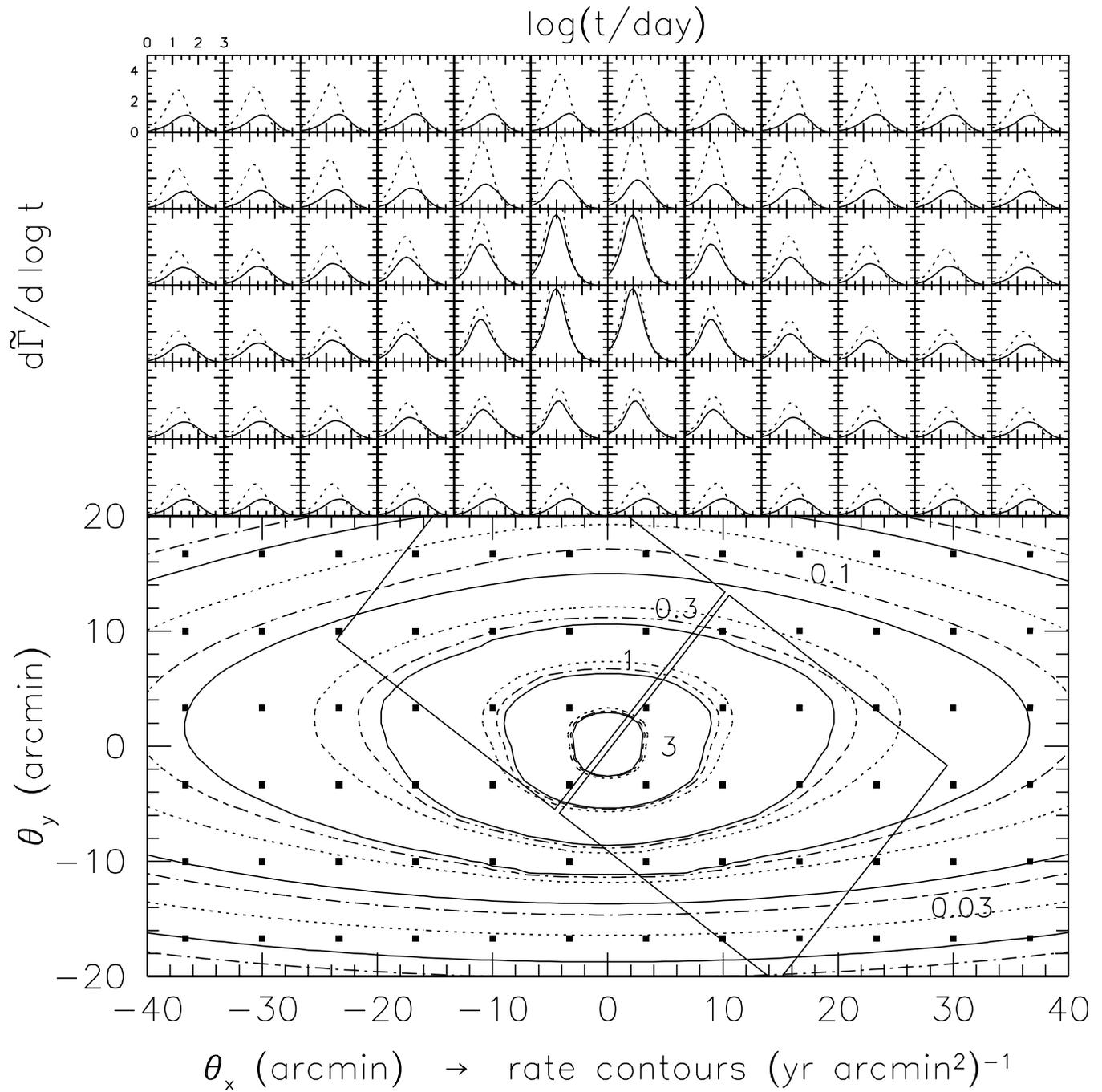,width=\columnwidth}
\caption{Rate contours for microlensing by lenses of different masses (in units
of events per year per square arcminute) and rate distributions for two
different halo lens masses.  Self lensing and Milky Way contributions are
included.  The bottom panel illustrates the $f_b=0.2$, $q=1$, $r_c=2$ kpc case,
with solid contours indicating a lens mass (for M31 and Milky Way halo
contributions) of $1.0 M_\odot$, dot-dashed contours indicating the fiducial
case $-3/8$ dex relative to solar, and the dotted contours indicating $0.1
M_\odot$ lenses.  As expected, the rate for the lowest--mass lenses is the
largest.  In the top panels $d\tilde{\Gamma}/d\log\thalf$ is plotted for the
$0.1 M_\odot$ (dotted curves) and $1.0 M_\odot$ (solid curves) cases, again
normalized to unity at $\log(\tthalf/{\rm day})=1.75$ (about 56 days).  In the
center there is little difference, as the self lensing dominates.  Away from
the bulge, the longer timescales and lower rates of the heavier lenses is
clearly seen.}
\label{fig:map_mass}
\end{figure}

Finally, we discuss the effects of varying the MACHO mass on the total and
differential rates.  The stellar mass function will remain that of Chabrier
(2001).  In the bottom panel of Fig.~\ref{fig:map_mass} rate contours are
plotted for fixed MACHO masses (for both the M31 and Milky Way contributions)
of $1.0 M_\odot$, $-3/8$ dex solar, and $0.1 M_\odot$.  As expected, the lower
masses give larger rates.  We have also given the MACHOs a Chabrier (2001)
function, identical to the stars.  The resulting rates are similar to the $0.1
M_\odot$ case (as the Chabrier mass function peaks there), thus we do not
illustrate them.  In the top panels of Fig.~\ref{fig:map_mass} we show
$d\tilde{\Gamma}/d\log\thalf$ for lens masses of $1.0 M_\odot$ and $0.1
M_\odot$.  The shift in peak is clearly seen, as is the higher rate for smaller
lenses.  Taking a Chabrier (2001) mass function, the rate distribution is very
similar to the fiducial case illustrated in Fig~\ref{fig:map_dist}.

\section{Discussion}
\label{discussion}

We have provided maps of the microlensing event rate towards M31 in a model
including M31 disk and bulge sources, and lenses from the M31 disk and bulge,
as well as lenses in the dark halos of M31 and the Milky Way.  We have varied
the parameters of the dark halo, namely the core radius and the flattening, and
studied the effects on the rate contours.  The core radius affects the rates
most in the galaxy center but these changes are partially obscured by the high
rate of self lensing events there.  The flattening does have a significant
effect, especially along the major axis.  The normalization of the rate
contours away from the bulge gives a measure of the halo MACHO fraction.

The easily measured timescale parameter is the full width at half maximum, but
the more physical timescale is the Einstein time.  Knowing the Einstein times
allows a much more accurate measurement of the lens mass, thus it is worth the
extra trouble to try to measure the Einstein times of the detected events.

We have shown that a microlensing halo in M31 should be clearly distinguishable
from self lensing if an appreciable event rate away from the M31 bulge is
measured.  We have explored the use of cuts in event timescale to separate the
self lensing component from the halo lensing component of the event rate, and
found that this helps near the bulge, but not further away.  We have quantified
the level of halo that is detectable, and found that a marginal detection of a
5\% microlensing halo would be possible in three seasons of ground--based
observations.  Higher halo fractions can be detected more convincingly of
course.

\acknowledgments

We wish to thank David Alves, Krzysztof Stanek and Larry Widrow for useful
conversations.  E.B. acknowledges support from the Columbia University Academic
Quality Fund.  G.G. wishes to acknowledge financial support from the Brinson
Foundation.  A.C. was supported by grants from NSF (AST 00-70882 and 98-02984)
and \mbox{STScI (GO-7376)}.

\end{document}